\documentclass[a4paper,12pt]{article}
\usepackage[pctex32]{graphics}
\usepackage{amssymb,amsmath}

\begin{document}
\topmargin 0pt
\oddsidemargin 0mm
\newcommand{\be}{\begin{equation}}
\newcommand{\ee}{\end{equation}}
\newcommand{\ba}{\begin{eqnarray}}
\newcommand{\ea}{\end{eqnarray}}
\newcommand{\fr}{\frac}
\renewcommand{\thefootnote}{\fnsymbol{footnote}}

\begin{titlepage}

\vspace{5mm}
\begin{center}
{\Large \bf Dilaton gravity approach to three dimensional Lifshitz
black hole }

\vskip .6cm
 \centerline{\large
 Yun Soo Myung$^{1,a}$, Yong-Wan Kim $^{1,b}$,
and Young-Jai Park$^{2,c}$}

\vskip .6cm

{$^{1}$Institute of Basic Science and School of Computer Aided
Science,
\\Inje University, Gimhae 621-749, Korea \\}

{$^{2}$Department of Physics and Center for Quantum Spacetime,\\
Sogang University, Seoul 121-742, Korea}
\end{center}

\begin{center}

\underline{Abstract}
\end{center}

 The $z=3$ Lifshitz black hole is an exact
black hole solution to the new massive gravity in three dimensions.
In order to understand this black hole clearly,  we  perform a
dimensional reduction to two dimensional dilaton gravity by
utilizing the circular symmetry. Considering the linear dilaton, we
find the same Lifshitz black hole in two dimensions. This implies
that all thermodynamic quantities of the $z=3$ Lifshitz black hole
could be obtained from its corresponding black hole in two
dimensions. As a result, we derive the temperature, mass, heat
capacity,  Bekenstein-Hawking entropy, and free energy.

\vspace{5mm}


\vskip 0.8cm

\vspace{15pt} \baselineskip=18pt
\noindent $^a$ysmyung@inje.ac.kr \\
\noindent $^b$ywkim65@gmail.com\\
\noindent $^c$yjpark@sogang.ac.kr

\thispagestyle{empty}
\end{titlepage}

\newpage
\section{Introduction}
Recently, the Lifshitz-type black
holes~\cite{CFT-4,L-1,AL-3,L-2,L-4,L-3,L-5} have received
considerable attentions since these may provide a model of
generalizing AdS/CFT correspondence to non-relativistic condensed
matter physics~\cite{CFT-1,CFT-2,CFT-3}. However, although their
asymptotic spacetimes  are apparently simple, the problem of
obtaining an analytic exact solution seems to be a highly nontrivial
task. A few examples include  a four-dimensional topological black
hole which is asymptotically Lifshitz with the dynamical exponent
$z=2$~\cite{Mann}. An analytic black hole solution with $z=2$ that
asymptotes the planar Lifshitz spacetime was found in
four-dimensional spacetimes~\cite{bm}, and the $z=3$ Lifshitz black
hole~\cite{z3} was derived from the new massive gravity (NMG) in
three-dimensional
spacetimes~\cite{bht,nmg-1,nmg-2,nmg-3,nmg-4,nmg-5,nmg-6}. Numerical
solutions were also explored in ~\cite{BBP,AL-2}. However, their
complete thermodynamic studies  are limited because it is not easy
to compute their conserved quantities in asymptotic Lifshitz.

On the other hand, two-dimensional (2D) dilaton gravity has been
used in various situations as an effective description of 4D and
3D gravities after a black hole in string theory has
appeared~\cite{wit,wit1}. It is known that the 2D dilation gravity
approach completely preserves the thermodynamics of 4D and 3D
black
holes~\cite{2D-1,2D-2,2D-3,Nojiri00,2D-4,JT-1,JT-2,JT-3,JT-4,DR-1,DR-2,DR-3,DR-4,DR-5,DR-6,DR-7}.

Hence, it is quite reasonable  to apply the 2D dilaton gravity
approach to the Lifshitz black holes in order to find their
thermodynamic quantities. In this work, first, we check that the
$z=3$ Lifshitz black hole is also a solution to the 2D dilaton
gravity. Then, {\it  we use the 2D dilaton gravity approach to the
$z=3$ Lifshitz black hole in three-dimensional spacetimes to obtain
all thermodynamic quantities.}  In addition, we wish to point out
differences and similarities between the $z=3$ Lifshitz black hole
and the $z=1$ nonrotating BTZ black hole in the 2D dilaton gravity
approach.

\section{3D New Massive Gravity}

The NMG action~\cite{bht} composed of the
Einstein-Hilbert action with a cosmological constant $\lambda$ and
higher order curvature terms is given by
\begin{eqnarray}
\label{NMGAct}
 S^{(3)}_{NMG} &=& S^{(3)}_{EH}+S^{(3)}_{HC}, \\
\label{NMGAct2} S^{(3)}_{EH} &=& \frac{1}{16\pi G_3} \int d^3x \sqrt{-\cal{G}}~ ({\cal R}-2\lambda),\\
\label{NMGAct3} S^{(3)}_{HC} &=& -\frac{1}{16\pi G_3m^2} \int d^3x
            \sqrt{-\cal{G}}~\left({\cal R}_{MN}{\cal R}^{MN}-\frac{3}{8}{\cal R}^2\right),
\end{eqnarray}
where $G_3$ is a three-dimensional Newton constant and $m^2$ a
parameter with mass dimension 2. From now on, we set $G_3 = 1/8$
to obtain the same normalization of the refs. \cite{L-5,AO}.

The field equation is given by \be {\cal
R}_{MN}-\frac{1}{2}g_{MN}{\cal R}+\lambda
g_{MN}-\frac{1}{2m^2}K_{MN}=0,\ee where
\begin{eqnarray}
  K_{MN}&=&2\square {\cal R}_{MN}-\frac{1}{2}\nabla_M \nabla_N {\cal R}-\frac{1}{2}\square{\cal R}g_{MN}\nonumber\\
        &+&4{\cal R}_{MNPQ}{\cal R}^{PQ} -\frac{3}{2}{\cal R}{\cal R}_{MN}-{\cal R}_{PQ}{\cal R}^{PQ}g_{MN}
         +\frac{3}{8}{\cal R}^2g_{MN}.
\end{eqnarray}
In order to have Lifshitz black hole solution with
dynamical exponent $z$, it is convenient to introduce dimensionless
parameters \be y=m^2~ \ell^2,~~w=\lambda~ \ell^2,
\ee
where $y$ and $w$ are proposed to take
\be
y=-\frac{z^2-3z+1}{2},~~w=-\frac{z^2+z+1}{2}.
\ee
For the $z=1$ nonrotating BTZ black hole, one has $y=\frac{1}{2}$ and
$w=-\frac{3}{2}$, while
 $y=-\frac{1}{2}$ and
$w=-\frac{13}{2}$ are chosen for $z=3$ Lifshitz black hole.

Now, let us consider the Achucarro-Ortiz type of dimensional
reduction \cite{AO} by introducing the dilaton $\Phi$ as
\begin{equation}
ds^2_{(3)}=g_{\mu\nu}(x)dx^\mu dx^\nu +\ell^2\Phi^2(x) d\theta^2.
\end{equation}
After integration over $\theta$ on $S^1$, the action (\ref{NMGAct}) is
reduced to give a 2D effective dilaton action
\begin{eqnarray}
\label{2dMGAct}
 S_{NMG} &=& S_{EH}+S_{HC}, \\
 S_{EH} &=& \ell \int d^2x \sqrt{-g}~ \Phi \Big(R-2\lambda\Big), \\
 S_{HC} &=& -\frac{\ell}{2m^2} \int d^2x \sqrt{-g}~ \Phi
         \left[\frac{1}{4}R^2+\frac{1}{\Phi}R\nabla^2\Phi
         +\frac{2}{\Phi^2}\nabla_\rho\nabla_\sigma\Phi\nabla^\rho\nabla^\sigma\Phi \right.\nonumber\\
     &&~~~~~~~~~~~~~~~~~~~~~~~~~~~~~~~~\left. -\frac{1}{\Phi^2}(\nabla^2\Phi)^2\right].
\end{eqnarray}
We note that higher derivatives containing ${\cal R}_{MN}{\cal
R}^{MN}$ and ${\cal R}^2$ are partially  realized into the dilaton
field. At this stage, we wish to distinguish 3D curvature ${\cal
R}$ from 2D curvature $R$. After a lengthy calculation, we obtain
the equations of motion for 2D metric tensor $g^{\mu\nu}$ and
dilaton $\Phi$ as
\begin{eqnarray}
\label{eomg}
  &&   \lambda \Phi g_{\mu\nu}+g_{\mu\nu}\nabla^2\Phi-\nabla_\mu\nabla_\nu\Phi \nonumber\\
  && -\frac{1}{2m^2}\left[-\frac{1}{2}g_{\mu\nu}\Phi\left(\frac{1}{4}R^2+\frac{1}{\Phi}R\nabla^2\Phi
     +\frac{2}{\Phi^2}\nabla_\rho\nabla_\sigma\Phi\nabla^\rho\nabla^\sigma\Phi-\frac{1}{\Phi^2}(\nabla^2\Phi)^2\right)\right.\nonumber\\
  &&~~~~~~~~~~ +\frac{1}{2}\Phi RR_{\mu\nu}+\frac{1}{2}g_{\mu\nu}\nabla^2(\Phi R)-\frac{1}{2}\nabla_\mu\nabla_\nu(\Phi R)
               +R_{\mu\nu}\nabla^2\Phi \nonumber\\
  &&~~~~~~~~~~ +g_{\mu\nu}\nabla^4\Phi-\nabla_\mu\nabla_\nu(\nabla^2\Phi)+R\nabla_\mu\nabla_\nu\Phi
               +2\nabla_\mu(R\nabla_\nu\Phi) \nonumber\\
  &&~~~~~~~~~~ -g_{\mu\nu}\nabla_\rho(R\nabla^\rho\Phi)
    +\frac{2}{\Phi}\Big(\nabla_\mu\nabla_\rho\Phi\nabla_\nu\nabla^\rho\Phi+\nabla_\rho\nabla_\mu\Phi\nabla^\rho\nabla_\nu\Phi\Big)\nonumber\\
  &&~~~~~~~~~~ +2\nabla_\rho\Big(\frac{1}{\Phi}\nabla_\mu\nabla_\nu\Phi\nabla^\rho\Phi
                             -\frac{2}{\Phi}\nabla^\rho\nabla_\mu\Phi\nabla_\nu\Phi\Big)
               -\frac{2}{\Phi}\nabla^2\Phi\nabla_\mu\nabla_\nu\Phi          \nonumber\\
  &&~~~~~~~~~~  \left. +2\nabla_\mu\Big(\frac{1}{\Phi}\nabla^2\Phi\nabla_\nu\Phi\Big)
                -g_{\mu\nu}\nabla_\rho\Big(\frac{1}{\Phi}\nabla^2\Phi\nabla^\rho\Phi\Big)\right]=0,
\end{eqnarray}
\begin{eqnarray}
\label{eomphi}
  && \Phi(R-2\lambda)-\frac{1}{2m^2}\left[\frac{1}{4}\Phi R^2
     +\Phi\nabla^2R-\frac{2}{\Phi}\nabla_\rho\nabla_\sigma\Phi\nabla^\rho\nabla^\sigma\Phi\right.     \nonumber\\
  && ~~~~~~\left.+4\Phi\nabla_\rho\nabla_\sigma\Big(\frac{1}{\Phi}\nabla^\rho\nabla^\sigma\Phi\Big)
             +\frac{1}{\Phi}\Big(\nabla^2\Phi\Big)^2-2\Phi\nabla^2\Big(\frac{1}{\Phi}\nabla^2\Phi\Big)\right]=0,
\end{eqnarray}
respectively. Moreover, the trace part of the equation of motion
(\ref{eomg}) is given by
\begin{eqnarray}
  && 2\lambda\Phi+\nabla^2\Phi
    -\frac{1}{2m^2}\left[\frac{1}{4}\Phi R^2+R\nabla^2\Phi+\nabla^4\Phi+\frac{1}{2}\nabla^2(\Phi R)\right. \nonumber\\
  && \left. +\frac{2}{\Phi}\nabla_\rho\nabla_\sigma\Phi\nabla^\rho\nabla^\sigma\Phi
      -\frac{1}{\Phi}(\nabla^2\Phi)^2
    -4\nabla_\rho\Big(\frac{1}{\Phi}\nabla_\sigma\Phi\nabla^\rho\nabla^\sigma\Phi\Big)
    +2\nabla_\rho\Big(\frac{1}{\Phi}\nabla^\rho\Phi\nabla^2\Phi\Big)\right]=0.     \nonumber\\
\end{eqnarray}
In deriving the above equations, we use $R_{\mu\nu}=Rg_{\mu\nu}/2$.

 Considering the linear dilaton background,
 \be
 \Phi=\frac{r}{\ell},\ee
 we find the $z=3$ Lifshitz black hole in two spacetimes
 \be
\label{2dmetric}
  ds^2_{(2),~z=3}=g_{\mu\nu}dx^\mu dx^\nu=-\left(\frac{r^2}{\ell^2}\right)^3\left(1-\frac{M\ell^2}{r^2}\right)dt^2
   +\frac{dr^2}{\left(\frac{r^2}{\ell^2}-M\right)},
\end{equation}
where $M$ is an integration constant related to the the ADM mass
of black hole. For $z=1$, the ADM mass is determined to be
$M=\frac{r_+^2}{\ell^2}$, while for $z=3$, the ADM mass is
proportional to $M^2$ as discussed in the next section.  The
curvature invariants of the $z=3$ Lifshitz black hole solutions in
two-dimensional spacetimes are given by
\begin{equation}
  R=-\frac{18}{\ell^2}+\frac{4M}{r^2},
  ~~~R_{\mu\nu}R^{\mu\nu}=\frac{R^2}{2}=\frac{162}{\ell^4}-\frac{72M}{r^2\ell^2}+\frac{8M^2}{r^4},
\end{equation}
which show the curvature singularity at the origin. We observe that
the singular behaviors of ${\cal R}$ and ${\cal R}_{MN}{\cal
R}^{MN}$~\cite{z3} persist in the 2D dilation gravity.

Furthermore, the above metric could be expressed in terms of the
dilaton $\Phi$ as \be \label{2pdmetric}
  ds^2_{(2),z}=-\Phi^{2z}\left(1-\frac{M\ell^2}{r^2}\right)dt^2
   +\frac{dr^2}{\left(\frac{r^2}{\ell^2}-M\right)}.
\end{equation}
For the $z=1$ case, which corresponds to the particular point $y = 1/2, w=
- 3/2$, we find  the nonrotating BTZ black hole, while for the $z=3$ case,
which corresponds to the particular point $y = - 1/2, w= - 13/2$,
Lifshitz black hole is recovered. However, the $z=2$ case is not a
solution to the 2D dilaton gravity (\ref{2dMGAct}).

We note that for the $z=1$ case, $M$ is the ADM mass because it could be
calculated in asymptotically AdS spacetimes using the Hamiltonian
formalism. However, for the $z=3$ case, we could not identify $M$ as the
ADM mass because it should be calculated in asymptotically
Lifshitz spacetimes. According to the information on
Ho\v{r}ava-Lifshitz black holes~\cite{CCO1,CCO2}, which are also
Lifshitz black holes with $0\le z\le 4$~\cite{MK}, it is
conjectured that the ADM mass ${\cal M}$ is given by \be
\label{masscon}
 M \propto \sqrt{{\cal M}}.
 \ee

\section{ Thermodynamics of $z=3$ Lifshitz black hole}
First of all, we mention that the Hawking temperature,
which is related with the Lifshitz black hole solution
(\ref{2dmetric}), can be determined from the metric as
\begin{equation} \label{temp}
 T_H =
 \frac{1}{4\pi}\Big[\sqrt{-g^{tt}g^{rr}}~ \left|g_{tt}'(r)\right|\Big]_{r=r_+}
     = \frac{r^3_+}{2\pi\ell^4},
\end{equation}
irrespective of knowing other conserved quantities. Here $^\prime$
denotes differentiation with respect to its argument. In Ref.
\cite{MKP1}, it is well known that all thermodynamic quantities of
the 4D Reissner-Nordstr\"om black hole and 3D BTZ black
hole~\cite{BTZ-1,BTZ-2} can be  expressed in terms of dilaton,
dilaton potential $V(\Phi)$, its integration $J(\Phi)$, and its
derivative $V'(\Phi)$ in its  2D dilaton gravity. Explicitly, their
corresponding relations of temperature $T_H$, mass $J$, and heat
capacity $C$ are given by as
\begin{equation} \label{potthe}
  T_H=\frac{V(\Phi)}{4\pi\ell},
  ~~J=\int^\Phi
  V(\tilde{\Phi})d\tilde{\Phi},~~C=4\pi\frac{V(\Phi)}{V'(\Phi)}.
\end{equation}
Therefore, expressing the temperature (\ref{temp}) as a function
of potential
\begin{equation}
T_H=\frac{V(\Phi)}{4\pi\ell},
\end{equation}
with potential
\begin{equation}
V(\Phi)=2\Phi^3=\frac{2r^3_+}{\ell^3},
\end{equation}
the mass $J$ and heat capacity $C$ are found to be
\begin{eqnarray} \label{mass}
 J(\Phi)
 &=&\frac{\Phi^4}{2}=\frac{r^4_+}{2\ell^4}, \\
\label{heatc}  C&=&\frac{4\pi}{3}\Phi=\frac{4\pi r_+}{3\ell},
\end{eqnarray}
respectively. Here we note that the mass $J=\Phi^4/2$ of $z=3$
Lifshitz black hole takes a different form, compared to the mass
$M=\Phi^2(r_+)=r_+^2/\ell^2$ of the nonrotating $z=1$ BTZ black
hole.  From the conjecture of (\ref{masscon}) inspired by the
Ho\v{r}ava-Lifshitz black holes, we expect that the ADM mass of
$z=3$ Lifshitz black hole is determined to  be   \be {\cal M}
\propto M^2=\frac{r_+^4}{\ell^4} \ee up to the constant.  Using the
dilaton gravity approach to the $z=3$ Lifshitz black hole, however,
we determine the ADM mass of $z=3$ Lifshitz black hole exactly as
\be {\cal M} \to J=\frac{r_+^4}{2\ell^4} \ee without calculating the
mass in asymptotically Lifshitz\footnote{Recently, a boundary
stress-tensor approach  has shown that the negative sign
Einstein-Hilbert term provides a consistent  thermodynamics of the
$z=3$ Lifshitz black hole obtained from the NMG~\cite{HT}, which was
the exactly same result derived here.}. Using the first law of
thermodynamics, \be dJ=T_HdS, \ee we derive the Bekenstein-Hawking
entropy
\begin{equation}\label{BHentropy}
 S=4\pi r_+,
\end{equation}
which satisfies the area-law for a universal entropy of black
holes\footnote{Note here that when considered the Newton's
constant, the mass $J$ in (\ref{potthe}) is expressed by
$J=\int^\Phi V(\tilde{\Phi})d\tilde{\Phi}/(8G_3)$, $M\rightarrow
8\pi G_3 M$, and the righthand sides of $S$, $F$ are divided by
$8G_3$, respectively.}. Finally, the free energy is given by \be
F=J-T_HS=-\frac{3}{2}\frac{r_+^4}{\ell^4}\to
-\frac{3}{2}\Phi^4.\ee

It seems appropriate to comment on the $z=1$ case.
For this case, the temperature is given by
\begin{equation}
 T_H(\Phi)=\frac{\Phi}{2\pi\ell}=\frac{r_+}{2\pi\ell^2},
\end{equation}
while the mass and specific heat take the forms
\begin{equation}
 J(\Phi)=\Phi^2=\frac{r_+^2}{\ell^2},~~~C(\Phi)=4\pi\Phi=\frac{4\pi r_+}{\ell}.
\end{equation}
Using the first law of thermodynamics, we also have the
Bekenstein-Hawking entropy
\begin{equation} \label{entropy}
 S=4\pi r_+.
\end{equation}
The free energy is given by~\cite{myungbtz}
\be
F=-\Phi^2=-\frac{r_+^2}{\ell^2}.
\ee

\begin{figure}[t!]
   \centering
   \includegraphics{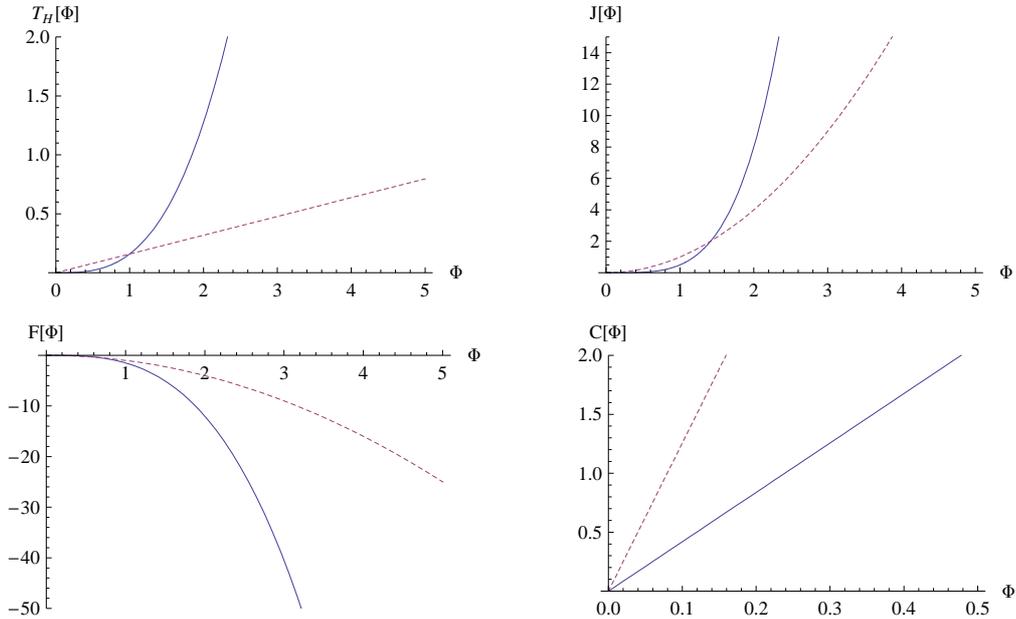}
\caption{Figures for temperature $T_H(\Phi)$, mass $J(\Phi)$, free
energy $F(\Phi)$, and specific heat $C(\Phi)$, and  with $\ell=1$:
Solid lines are for $z=3$, while dotted lines are for $z=1$.}
\label{fig.1}
\end{figure}
 Before we proceed, we would like to mention
that the key observation is the temperature expression (\ref{temp})
and thus, we have derived thermodynamic quantities based on  the
dilaton gravity approach (\ref{potthe}). This approach works well
for the BTZ and Reissner-Nordstr\"om black holes without higher
curvature terms. Here the higher curvature terms (\ref{NMGAct3})
appeared and these are necessary to obtain the $z=3$ Lifshitz black
hole like the warped AdS$_3$ solution in the topologically massive
gravity~\cite{ALPSS}. If we know Lifshitz asymptotes well, we may
calculate conserved quantities at infinity using the Hamiltonian
formalism. However, at this time, one does not know precisely how
Lifshitz asymptotes is different from asymptotically A(dS)
spacetimes. Hence, we have an intrinsic  handicap to determine the
thermodynamic quantities of $z=3$ Lifshitz black holes using the
conventional approach. Therefore, one has to find an alternative,
even it was not proved to be working for the $z=3$ Lifshitz black
hole well.

In this work, we might provide one way to determine  thermodynamic
quantities of $z=3$ Lifshitz black holes by using the 2D dilaton
gravity approach. This attempt was supported from the fact that
all higher dimensional black holes could be obtained from the 2D
dilaton gravity approach after making an appropriate dimensional
reduction. In this case, it is important to compare our results
with the known results.

At this stage, we compare our thermodynamic quantities with those
of ref.\cite{L-5}.
 It seems that the temperature (\ref{temp}) is
the same, but mass (\ref{mass}) and entropy (\ref{entropy}) are
consistent with those in \cite{L-5} except  negative signs when
choosing  $L=2\pi \ell$. As was explained in  \cite{L-5}, negative
mass and entropy are unfamiliar to black hole physicists and thus,
this problem  may be resolved when replacing the Newton's constant
$G_3$ by $-G_3$. We insist that this replacement is indeed necessary
to find the NMG as  a unitary massive gravity~\cite{NO}. The NMG is
equivalent to the Fierz-Pauli massive gravity within the linearized
theory. In three dimensions, a massless graviton has no propagating
degrees of freedom, while a massive graviton is a physically
propagating mode with two helicities. In constructing the NMG with
higher curvature terms, the principle was that one can  neglect the
massless graviton from (\ref{NMGAct2}) whatever its norm is positive
or negative, in favor of the massive graviton without ghost from
(\ref{NMGAct3})~\cite{bht}. In this sense, the replacement of
$G_3\to -G_3$ is necessary to obtain the correct thermodynamic
quantities. In our work, this replacement should be done on the
action (\ref{NMGAct}) to obtain a correct action for a unitary
massive gravity. However, this global operation does not change our
thermodynamic results because we have derived thermodynamic
quantities using the 2D dilaton gravity approach (mainly used the
equations of motion) but not the effective action to derive the
entropy by Wald's formula. The latter is sensitive to sign of Ricci
scalar and thus, leading to negative entropy and mass using the
first law of thermodynamics.  Recently, a boundary stress-tensor
approach  has confirmed  that the wrong (negative) sign
Einstein-Hilbert term provides a consistent  thermodynamics of the
$z=3$ Lifshitz black hole obtained from the NMG~\cite{HT}, which was
the exactly same result found in our work.

Consequently, there is no difference between our thermodynamic
quantities and those of \cite{L-5} if one  considers the NMG for a
unitary massive gravity seriously. In this case, we have still
found familiar thermodynamic quantities even for $z=3$ Lifshitz
black hole in three dimensions.

\section{Discussions}
First of all, we have derived all thermodynamic quantities of the $z=3$
Lifshitz black hole in three-dimensional spacetimes according to the
dilaton gravity approach. This suggests that unknown  thermodynamic
quantities of higher-dimensional Lifshitz black holes could be
obtained when using their 2D dilaton gravity approaches.

Next, we would like to mention differences and similarities between
the $z=3$ Lifshitz black hole and the $z=1$ nonrotating BTZ black hole in
the 2D dilaton gravity approach. As is shown Fig. 1, temperature,
mass, and free energy take different forms as \be T_H \propto r_+^z,~~
J \propto r_+^{z+1},~~F \propto -r_+^{z+1},\ee while the heat
capacity takes the nearly same forms as \be C \propto r_+. \ee
Importantly, the entropy is exactly the same  for $z=1,3$ black
holes.

At this stage, we would like to mention the stability issue on the
relation between 2D dilaton black hole and $z=3$ Lifshitz black
holes. This issue on the rotating BTZ black hole was discussed in
~\cite{ON1,ON2}, showing that taking into account quantum
corrections may lead to some instability. Also, some reduction to
low dimensions may spoil the equivalence between higher dimensional
and lower dimensional objects~\cite{DR-2}.  It is interesting to
investigate the stability issue on the the relation between 2D
dilaton black hole and $z=3$ Lifshitz black holes. However, we
remind the reader that at this time, one does not know precisely how
Lifshitz asymptotes is different from asymptotically A(dS)
spacetimes.  Hence, we have some difficulty  to study the stability
issue. Further, we could not see  whether the equivalence between
higher dimensional and lower dimensional objects is spoiled for
objects in the Lifshitz spacetimes. An important thing is that the
dimensional reduction will not change thermodynamic properties of
black holes. Therefore, we have shown that  thermodynamics of the 2D
dilaton black hole is the same as that of the 3D Lifshitz black
hole.

Consequently, it is strongly suggested  the 2D dilaton gravity
approach may shed light on studying thermodynamic properties of the
Lifshitz-type black holes.

\vskip 0.5cm

\section*{Acknowledgement}
Two of us (Y. S. Myung and Y.-J. Park) were supported by the
National Research Foundation of Korea (NRF) grant funded by the
Korea government (MEST) through the Center for Quantum Spacetime
(CQUeST) of Sogang University with grant number 2005-0049409.
Y.-W. Kim was supported by the Korea Research Foundation Grant
funded by Korea Government (MOEHRD): KRF-2007-359-C00007. Y.-J.
Park was also partially supported by the Korea Science and
Engineering Foundation (KOSEF) grant funded by the Korea
government (MEST) through WCU Program (No. R31-20002).

\end{document}